\documentclass[aps,pra,showpacs,notitlepage]{revtex4-1}

\usepackage{graphicx}

\begin{document}

\title{How to put quantum particles on magic bullet trajectories that can hit two targets without a clear line-of-sight}

\author{Holger F. Hofmann}
\email{hofmann@hiroshima-u.ac.jp}
\affiliation{
Graduate School of Advanced Sciences of Matter, Hiroshima University,
Kagamiyama 1-3-1, Higashi Hiroshima 739-8530, Japan
}

\begin{abstract}
Quantum particles move in strange ways, even when they propagate freely in space. As a result of the uncertainty principle, it is not possible to control the initial conditions of particle emission in such a way that the particle will definitely pass through two precisely defined positions along its path, even if it is possible to line up the two positions with the emitter. However, there is also an upside to the quantum mechanical laws of motion: constructive quantum interferences can actually raise probabilities to values higher than those permitted by classical causality. Here, it is shown that conventional interferometric methods can be used to prepare photons in a quantum state in which a non-vanishing fraction of particles will hit both of two possible targets, even though the direct line-of-sight connecting the two targets is blocked at the source. The demonstration of the effect is complicated by the uncertainty principle because the physical detection of a particle at one target disturbs the motion of the particle, making it impossible to determine whether the initial state of motion would have allowed the particle to hit the other target or not. It is nonetheless possible to determine the minimal fraction of ``magic bullet'' particles that must have hit both targets by showing that the number of particles hitting target $A$ is larger than the number of particles missing target $B$. Quantum interference effects can thus be used to optimize the path of particles in free space beyond the classical limit of motion along a straight line.  
\end{abstract}

\maketitle
The oddity of quantum mechanics rests on the introduction of mysterious ``superpositions'' of clearly distinct alternatives. With regard to the motion of a quantum particle, this is often represented by a superposition of complete paths, giving the impression that Newtonian laws of motion do not constrain the motion of quantum particles at all \cite{Saw14,Mag16,Zho17}. Alternatively, the same physics can be described by the propagation of probability waves, where interference effects result in state-dependent modifications to motion along a straight line \cite{Bra94,War02,Pfi04,Ber10,Koc11,Sch13,Yea13,Hal14}. It might come as a surprise that the physical nature of these modifications to Newton's laws is such a controversial issue, but experimental resolutions of these fundamental questions are undermined by the uncertainty principle, which limits the evidence to statistical distributions of outcomes with insufficient information on the path taken by each individual particle. It is therefore an important scientific challenge to identify the experimentally observable evidence for deviations from classical laws of motion in the available statistics obtained from measurements of a large number of particles. 

Recent advances in the control and characterization of quantum states have provided us with the means to measure the coherences of quantum states in space in time, as demonstrated in the groundbreaking work of Lundeen and coworkers \cite{Lun11}. A key to these new methods of control is the possibility of quantum superpositions of states that refer to different properties of the particle and therefore do not exclude each other \cite{Hof12,The18}. Optical systems provide an ideal platform for the realization of this extraordinary level of control, since the available techniques of beam shaping and interferometry allow us to control the spatiotemporal quantum states of individual photons with a precision that is difficult to achieve in other systems. In particular, it is possible to combine diffraction and interferometry to realize superpositions of states localized at a given position with states limited to a well-defined velocity in such a way that the statistical control of position and momentum can exceed the limit set by quantitative evaluations of the corresponding uncertainties \cite{Hof17,Hof18}. This new method of control can thus help us to uncover aspects of the quantum mechanical laws of motion that are usually hidden by the statistical noise associated with these uncertainty limits. 

The present paper introduces a particularly striking illustration of this effect, which I refer to as a quantum magic bullet. This reference is inspired by the concept of a magical bullet that hits its intended target no matter how it is aimed. If there are two targets, $A$ and $B$, a quantum particle can be aimed at both targets at once if the areas of the two targets are lined up along the path of propagation. It should be noted that only one of the two targets can actually be inserted in the path of the beam, since a physical target would block or diffract the path of the particles, invalidating the concept of ``aiming'' a particle at the second target. The evidence that the particles hit both of the targets is given by the separate probabilities of hitting each of the two targets. If both targets can be hit with a probability of one, it should be obvious that the particles all went through the target areas even if no physical target was inserted. In the opposite limit, one could aim every particle either at $A$ or at $B$, such that a shot at $A$ will definitely miss $B$ and vice versa. In that case, the sum of the probabilities $P(A)$ and $P(B)$ of hitting the respective targets cannot exceed a total probability of one. Any probability in excess of one is evidence for particles that hit both targets at the same time.

If particles move in straight lines, the initial position at which the quantum particle is emitted must line up with the two targets. It is therefore possible to identify initial positions with a clear line-of-sight connecting both targets with the source of the quantum particles. If these initial positions are blocked by a physical obstacle, it is impossible to line up both targets for a clear shot. For classical particles, this means that particles that move through target area $A$ will miss target area $B$ and vice versa. For quantum particles, interference effects can be used to modify the free space motion of particles \cite{Hof17,Hof18}. In the following, it is shown that a specific optimized interference of very different wavefunctions results in a certain percentage of particles that can hit both targets even though the initial positions with joint lines-of-sight are completely blocked. The effect reported here thus demonstrates that quantum interference effects can be used to enhance the ``aim'' of quantum particles beyond the limit that would apply to classical particles. 

\begin{figure}[th]
\begin{picture}(500,360)
\put(50,0){\makebox(360,360){
\scalebox{0.8}[0.8]{
\includegraphics{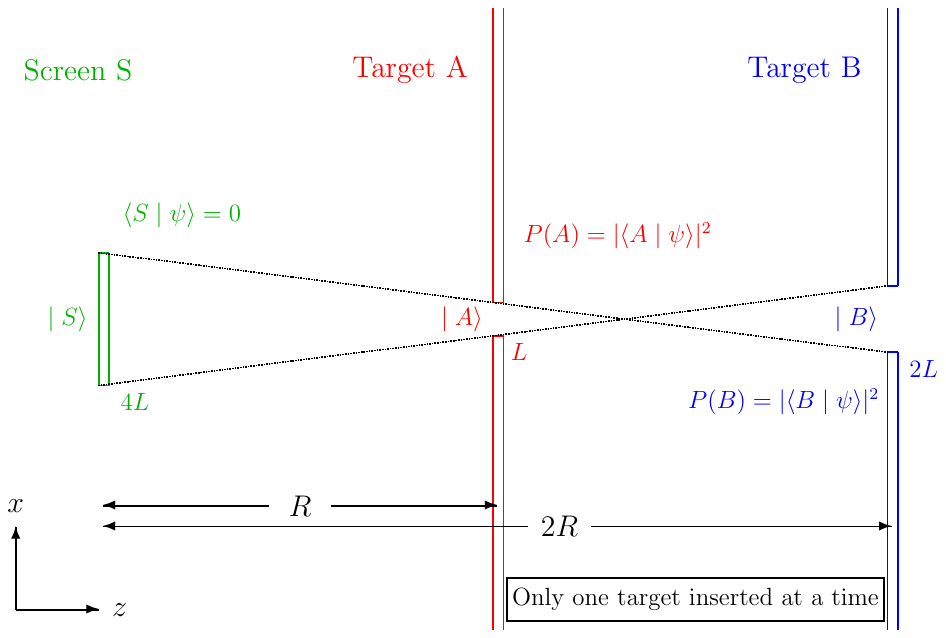}}}}
\end{picture}
\vspace{-3cm}
\caption{\label{fig1}
Possible optical setup for an experiment to verify the presence of quantum magic bullets in the state $\mid \psi \rangle$ defining the probability densities of transverse position $x$ along the axis of propagation $z$. Note that only one of the two targets should be inserted in each measurement, since the interaction between target $A$ and the particle will change the state and hence the transverse probability densities observed at the second target $B$. The thin black lines indicate the extremal lines-of-sight through both targets, showing that the screen at $S$ blocks all lines-of-sight. If a state satisfies $\langle S \mid \psi \rangle=0$, particles hitting target $A$ should not hit target $B$ and vice versa. However, quantum interference can result in a violation of this limit by up to 7.4 \% of the photons in $\mid \psi \rangle$.
}
\end{figure}

The precise physical scenario can be realized in a basic optical setup, where photons propagate along a single axis with an uncertainty limited transverse momentum describing the focus of the beam and its diffraction as it propagates along the axis. Fig. \ref{fig1} shows the geometry of the setup with the two possible targets $A$ and $B$ shown as apertures in two screens at distances of $R$ and $2R$ from a third screen $S$ that blocks all lines-of-sight through the apertures $A$ and $B$. Note that the two targets will never be inserted at the same time, since they will change the state of the photon by a projection on the state $\mid A \rangle$ or $\mid B \rangle$, rendering sequential measurements meaningless. The targets are only shown jointly to illustrate the blockage of line-of-sight propagation by the screen $S$. The slit widths of $L$ for the first target $A$ at $R$ and of $2L$ for the second target $B$ at $2 R$ from the screen at $S$ ensure that the wavefunctions of a beam $\mid A \rangle$ focused on target $A$ and the beam $\mid B \rangle$ focused on target $B$ have the same intensity profile at $S$ (see the appendix for the precise wavefunctions). The goal is to maximize the probability of hitting the targets by choosing an appropriate input state $\mid \psi \rangle$. This can be achieved by a superposition of the state $\mid A \rangle$ hitting target $A$ with a probability of $P(A)=1$, and the state $\mid B \rangle$ hitting target $B$ with a probability of $P(B)=1$. Each of the states corresponds to a wavefunction with a constant amplitude throughout the extent of the target. The phase of the states is chosen so that constructive interference results in an enhancement of the minimal probability of hitting both targets. As shown in the appendix, the wavefunctions of the quantum state components $\mid A \rangle$ and $\mid B \rangle$ at the different positions along the axis of propagation are fully determined by the requirement that they must be focused on their respective targets. Finally, the line-of-sight can be blocked by removing the state $\mid S \rangle$ representing the interval of width $4L$ at the source associated with all possible straight lines through both of the targets. It may be important to note that the interferometric subtraction of a beam component is also described by a propagation pattern, corresponding to a destructive interference with a third beam, the profile of which is determined by the focus on the interval covered by the screen at $S$. In principle, it is therefore possible to realize the state by interfering three different beam profiles. However, it is more simple and much easier to interpret the results if the removal of the $\mid S \rangle$ component of the state is realized by an actual physical obstruction, as suggested in Fig. \ref{fig1}. The resulting superposition of the three components is given by the single photon quantum state
\\
\begin{equation}
\label{eq:state}
\mid \psi \rangle = \frac{1}{\sqrt{2+2\langle A \mid B \rangle - \sigma^2}} \left( \mid A \rangle + \mid B \rangle - \sigma \mid S \rangle \right),
\end{equation}
\\
where the coefficient $\sigma$ is chosen so that the probability of finding the photons in the line-of-sight interval $S$ is exactly zero. It might be worth noting that the normalization of this state takes into account the overlap of the non-orthogonal states $\mid A \rangle$ and $\mid B \rangle$, as well as the condition that the overlap of the state $\mid \psi \rangle$ with the state $\mid S \rangle$ is zero. As can be confirmed, the normalization given ensures that $\langle \psi \mid \psi \rangle=1$, so that all measurement probabilities add up to a total probability of one. 

As shown in the appendix, the statistics of the state $\mid \psi \rangle$ are fully determined by the coefficient $\langle A \mid B \rangle$ describing the overlap between the beams focused on target $A$ and on target $B$, respectively. In turn, this coefficient is determined by the ratio of target size $L$ and target distance $R$, and the axial momentum $p_z$ of the quantum particles,
\\
\begin{equation}
\langle A \mid B \rangle = \sqrt{\frac{p_z L^2}{\pi \hbar R}}.
\end{equation}
\\
The location of Planck's constant in the denominator indicates that high values of the overlap correspond to the classical limit of line-of-sight propagation. As a result, we can expect that all photons propagating through both targets, $A$ and $B$, will be blocked at $S$ for sufficiently large overlap values $\langle A \mid B \rangle$. On the other hand, some overlap is necessary to overcome the effects of uncertainty which make it difficult to hit both targets in the extreme quantum limit. The maximal probabilities of hitting the targets are obtained between these two limits.

\begin{figure}[th]
\begin{picture}(500,360)
\put(50,0){\makebox(360,360){
\scalebox{0.8}[0.8]{
\includegraphics{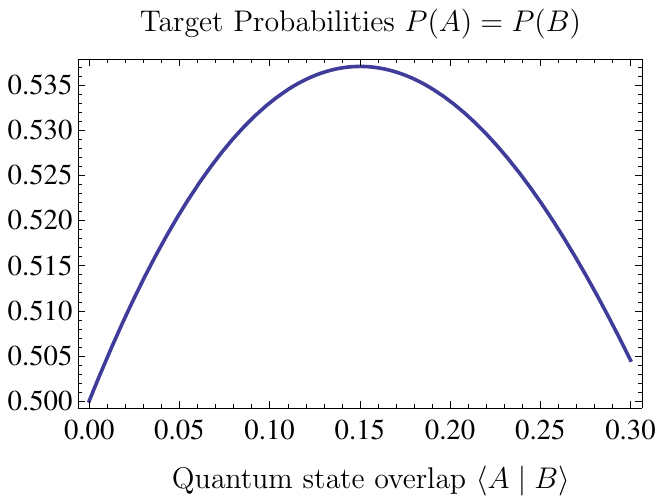}}}}
\end{picture}
\vspace{-4.5cm}
\caption{\label{fig2}
Dependence of the probabilities $P(A)$ and $P(B)$ to hit the respective targets on the quantum overlap $\langle A \mid B \rangle$ of the states associated with the targets. Quantum interference increases the target probabilities beyond 50\%. However, the amount of increase is limited by the increasing losses caused by the destructive interference with $\mid S \rangle$ which represents the obstruction of line-of-sight paths. A maximum of $P(A)=P(B)=0.537$ is reached at $\langle A \mid B \rangle=0.150$, corresponding to a minimal fraction of 7.4 \% of photons hitting both targets even though the lines-of-sight are blocked at $S$.  
}
\end{figure}

\begin{figure}[th]
\begin{picture}(480,420)
\put(125,375){\makebox(30,30){\Large (a)}}
\put(120,200){\makebox(240,200){\hspace{-1.5cm}\vspace{-2.5cm}
\scalebox{0.7}[0.7]{
\includegraphics{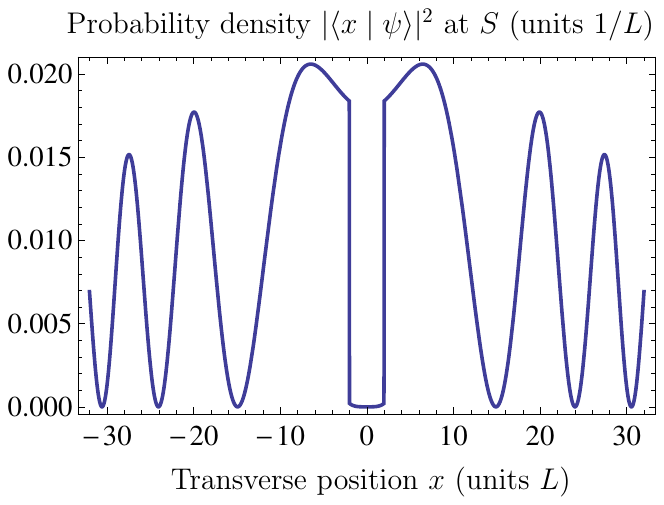}}}}

\put(25,180){\makebox(30,30){\Large (b)}}
\put(0,0){\makebox(240,200){\hspace{-1.5cm}\vspace{-2.5cm}
\scalebox{0.7}[0.7]{
\includegraphics{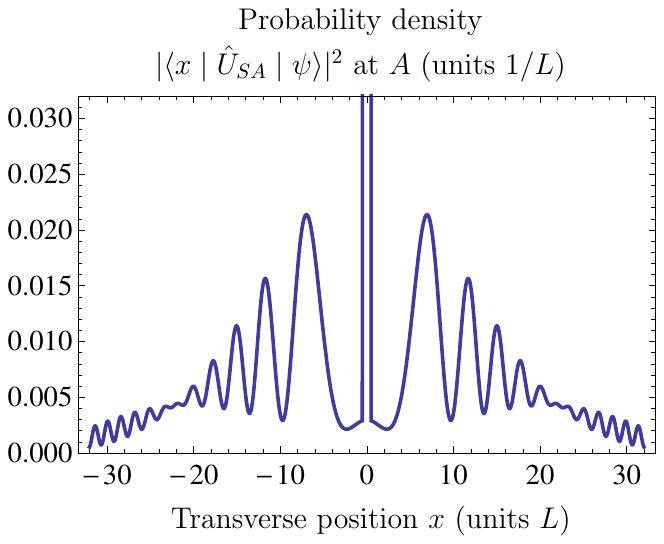}}}}

\put(265,180){\makebox(30,30){\Large (c)}}
\put(240,0){\makebox(240,200){\hspace{-1.5cm}\vspace{-2.5cm}
\scalebox{0.7}[0.7]{
\includegraphics{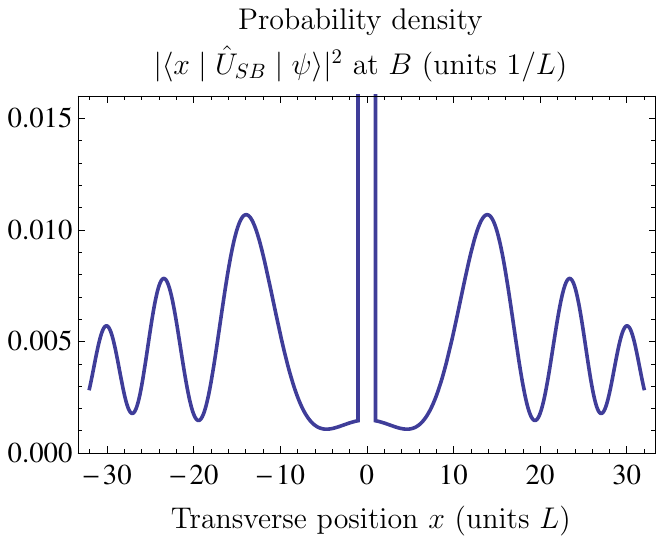}}}}

\end{picture}
\vspace{-0.5cm}
\caption{\label{fig3} Probability densities of particle detections for the optimized state $\mid \psi \rangle$ with a quantum overlap of $\langle A \mid B \rangle=0.150$. (a) shows the probability density at $S$, where the screen blocks out the line-of-sight positions at $|x|\leq 2L$. Quantum interferences enhancing the probabilities of hitting the targets $A$ and $B$ surround the blocked region. (b) shows the probability density at the target $A$. The central peak has a width of $L$ and a height of $0.537/L$. The interference pattern surrounding the central peak originates from the destructive interference between $\mid B \rangle$ and $\mid S \rangle$. The total probability of missing the target at $A$ is 46.3 \%. (c) shows the probability density at the target $B$. The pattern is the same as the one at $A$, but it is exactly twice as wide and half as dense. The central peak has a width of $2L$ and a height of $0.269/L$. Even if all of the 46.3 \% of photons that miss target $A$ were hitting target $B$ instead, at least 7.4 \% of the photons must be hitting target $A$ as well.  
}
\end{figure}

Fig. \ref{fig2} shows the dependence of the probabilities $P(A)$ and $P(B)$ on the overlap $\langle A \mid B \rangle$. The result can be approximately described by a linear increase in probability caused by the reduction of quantum uncertainty, and a quadratic decrease of probability caused by the destructive interference representing the blocked line-of-sight propagation at $S$,
\begin{equation}
P(A)=P(B) \approx \frac{1}{2} + \frac{1}{2} \langle A \mid B \rangle - (1+\frac{1}{\sqrt{2}}) \langle A \mid B \rangle^2.
\end{equation}
The precise calculation given in the appendix shows that the maximal probability is obtained at $\langle A \mid B \rangle = 0.150$, where the probabilities are $P(A)=P(B)=0.537$, or 53.7 \% of all photons. Even though it is not possible to detect the same photon twice, the fact that each of the targets are hit by more than half of the photons indicates that some of the photons must be hitting both targets. The minimal number of photons that must be hitting both targets can be found by subtracting the number of photons that miss target $B$ from the number of photons that hit target $A$ or vice versa. In terms of probabilities, 
\begin{equation}
P_{MB}=P(A)+P(B)-1.
\end{equation}
This is the minimal fraction of photons that must be hitting both targets even though there is no line-of-sight connecting the source with the two targets. Since this enhancement of accuracy is reminiscent of the properties of ``magic bullets'' of folklore, the probability $P_{MB}$ can be identified as the fraction of magic bullet photons in the initial state $\mid \psi \rangle$. Magic bullet photons exist in a regime where constructive quantum interference enhances the probability of hitting the targets more than the destructive interference representing the blocked lines-of-sight at $S$ diminishes it. The maximal fraction of magic bullets is achieved at a specific quantum overlap of $\langle A \mid B \rangle = 0.150$, where the fraction of magic bullet photons is $P_{MB} = 0.074$, or 7.4 \%. Fig. \ref{fig3} shows the beam profiles of the optimized quantum state at the source and at the two targets. The probability distribution at $S$ shows the blocked lines-of-sight connecting $A$ and $B$ at the center and the interference pattern that results in the enhancement of target probability outside of the blocked region. At the targets, a central peak of 53.7 \% probability is surrounded by a much wider background of 46.3 \% of the photons showing a complicated interference pattern between the state of the other target and the blocked state $\mid S \rangle$. Note that the magic bullet probability $P_{MB}$ of 7.4 \% is obtained under the assumption that all of the 46.3 \% of photons that missed target $A$ would have hit target $B$, and vice versa. 

In conclusion, it is possible to prepare a beam of quantum particles in such a way that a minimum of 7.4 \% of the particles must pass through both target areas, $A$ and $B$, even though the probability of finding them at a source point $S$ with a clear line-of-sight connecting both targets is zero. Although it is impossible to detect the actual path taken by a single photon, the statistics taken separately at $A$ and at $B$ are sufficient to reveal the presence of magic bullet photons in the state $\mid \psi \rangle$. The present scenario thus demonstrates that it is possible to control free space motion of quantum particles to an extent that would not be possible if classical laws of motion were applicable. This is particularly noteworthy since quantum uncertainty usually has the opposite effect of preventing levels of control that would be entirely possible in classical physics. The novel application of quantum interference effects proposed in this paper therefore points the way to a more efficient use of the hidden advantages of quantum coherence. 

This work has been supported by JST-CREST (JPMJCR1674), Japan Science and Technology Agency.

\vfill

\begin{appendix}
\section{States and wavefunctions}

As mentioned in the main body of the paper, the targets $A$ and $B$ and the blocked region at the source $S$ are all represented by rectangular wavefunctions in their respective focal planes. The evolution of these wavefunctions along the axis of propagation $z$ is then described by Fresnel diffraction. In the present case, it is possible to describe the results by appropriate sinc functions with a curved wavefront. Specifically, the wavefunctions of the components at the source $S$ are given by
\begin{eqnarray}
\langle x \mid S \rangle &=& \frac{1}{\sqrt{4L}} \; \mbox{for} \; |x|\leq 2L, \;\; 0 \;\; \mbox{else},
\nonumber \\
\langle x \mid A \rangle &=& \sqrt{\frac{2 L}{\langle A \mid B \rangle^2}}\; \frac{1}{\pi\, x}\; \sin\left( \frac{\langle A \mid B \rangle^2}{2 L} \pi\, x \right) \exp\left(- i \pi \left(\frac{\langle A \mid B \rangle^2}{2 L^2} x^2 - \frac{1}{8}\right)\right),
\nonumber \\
\langle x \mid B \rangle &=& \sqrt{\frac{2 L}{\langle A \mid B \rangle^2}}\; \frac{1}{\pi\, x}\; \sin\left( \frac{\langle A \mid B \rangle^2}{2 L} \pi\, x\right) \exp\left(- i \pi \left(\frac{\langle A \mid B \rangle^2}{4 L^2} x^2 + \frac{1}{8}\right)\right).
\end{eqnarray}
The corresponding wavefunctions at the first target $A$ are given by
\begin{eqnarray}
\langle x \mid \hat{U}_{SA} \mid S \rangle &=& \sqrt{\frac{L}{2 \langle A \mid B \rangle^2}}\; \frac{1}{\pi\, x}\; \sin\left( \frac{2 \langle A \mid B \rangle^2}{L} \pi\, x \right) \exp\left(i \pi \left(\frac{\langle A \mid B \rangle^2}{2 L^2} x^2 - \frac{1}{4}\right)\right),
\nonumber \\
\langle x \mid \hat{U}_{SA} \mid A \rangle &=& \frac{1}{\sqrt{L}} \exp(-i \frac{\pi}{8}) \; \mbox{for} \; |x|\leq L/2, \;\; 0 \;\; \mbox{else},
\nonumber \\
\langle x \mid \hat{U}_{SA} \mid B \rangle &=& \sqrt{\frac{L}{\langle A \mid B \rangle^2}}\; \frac{1}{\pi\, x}\; \sin\left( \frac{\langle A \mid B \rangle^2}{L} \pi\, x\right) \exp\left(- i \pi \left(\frac{\langle A \mid B \rangle^2}{2 L^2} x^2 + \frac{1}{8}\right)\right).
\end{eqnarray}
and the corresponding wavefunctions at the second target $B$ are given by
\begin{eqnarray}
\langle x \mid \hat{U}_{SB} \mid S \rangle &=& \sqrt{\frac{L}{\langle A \mid B \rangle^2}}\; \frac{1}{\pi\, x}\; \sin\left( \frac{\langle A \mid B \rangle^2}{L} \pi\, x\right) \exp\left(i \pi \left(\frac{\langle A \mid B \rangle^2}{4 L^2} x^2 - \frac{1}{4}\right)\right),
\nonumber \\
\langle x \mid \hat{U}_{SB} \mid A \rangle &=& \sqrt{\frac{2 L}{\langle A \mid B \rangle^2}}\; \frac{1}{\pi\, x}\; \sin\left( \frac{\langle A \mid B \rangle^2}{2 L} \pi\, x \right) \exp\left(i \pi \left(\frac{\langle A \mid B \rangle^2}{2 L^2} x^2 - \frac{3}{8}\right)\right),
\nonumber \\
\langle x \mid \hat{U}_{SB} \mid B \rangle &=& \frac{1}{\sqrt{2L}}  \exp(-i \frac{3 \pi}{8}) \; \mbox{for} \; |x|\leq L, \;\; 0 \;\; \mbox{else}.
\end{eqnarray}
For the purpose of calculating the probabilities $P(A)$ and $P(B)$, it is sufficient to know the overlaps between the wavefunctions. When one of the wavefunctions is rectangular, the overlap can be estimated by multiplying the amplitude at $x=0$ with the square root of the width of the rectangle. The results read
\begin{eqnarray}
\langle A \mid S \rangle &\approx& \sqrt{2} \langle A \mid B \rangle \exp(- i \frac{\pi}{8}) 
\nonumber \\
\langle B \mid S \rangle &\approx& \sqrt{2} \langle A \mid B \rangle \exp( i \frac{\pi}{8}) 
\end{eqnarray}
It is therefore possible to express all of the overlaps in terms of $\langle A \mid B \rangle$. In particular, the coefficient $\sigma$ that reduces the probability $P(S)$ of finding the particles in positions with a clear line-of-sight to both targets to zero is given by
\begin{equation}
\sigma = 2 \sqrt{2} \langle A \mid B \rangle \cos(\frac{\pi}{8}). 
\end{equation}  
The interferences of the wavefunctions described by $\mid \psi \rangle$ at an overlap of $\langle A \mid B \rangle=0.1502$ are shown in Fig. 3 of the paper. 

\vfill

\section{Optimization of target probabilities}

Using the overlaps between the quantum states given above, it is possible to determine the probabilities $P(A)=P(B)$ as a function of $\langle A \mid B \rangle$. The probability of hitting target $A$ is given by
\begin{eqnarray}
|\langle A \mid \psi \rangle|^2 &=& \frac{|1+\langle A \mid B \rangle-\sigma \langle A\mid S \rangle|^2}{2+2\langle A \mid B \rangle - \sigma^2}
\nonumber \\
&=& \frac{1}{2} + \frac{1}{2} \langle A \mid B \rangle - (1+\frac{1}{\sqrt{2}}) \langle A \mid B \rangle^2 + \frac{\langle A \mid B \rangle^4}{1+\langle A \mid B \rangle-(2+\sqrt{2}) \langle A \mid B \rangle^2}.
\end{eqnarray}
Note that this probability is slightly higher than the approximate formula given in the main text, but only by a contribution of fourth order in $\langle A \mid B \rangle$. At the maximum ($\langle A \mid B \rangle=0.15$), the fourth order correction is about $5 \times 10^{-4}$ or one twentieth of a percent. 

The complete formula for the minimal fraction of magic bullet photons in the state $\mid \psi \rangle$ is
\begin{equation}
P_{MB} = P(A)+P(B)-1 = \langle A \mid B \rangle - (2+\sqrt{2}) \langle A \mid B \rangle^2 + \frac{2 \langle A \mid B \rangle^4}{1+\langle A \mid B \rangle-(2+\sqrt{2}) \langle A \mid B \rangle^2}.
\end{equation}
In this formula, the fourth order correction adds 0.1 \% to the percentage of magic bullet photons in the state. Using the complete formulas including the fourth order correction terms, the more precise calculation results for the maximal probabilities to hit the targets $A$ and $B$ are $P(A)=P(B)=0.537062$ at $\langle A \mid B \rangle=0.150$, corresponding to a minimal fraction of magic bullet photons of $P_{MB}=0.074124$, which is slightly higher than the rounded value of 7.4 \% given in the main text. The approximate formula would result in a maximal value of only 7.3 \%, as expected from the magnitude of the correction near $\langle A \mid B \rangle=0.150$.

\end{appendix}

\end{document}